\documentclass[a4paper,twoside]{article}

\usepackage{cite}
\usepackage{amsmath}

\input ijmpblat

\begin{document}
\runninghead{T.~Mishonov \textit{et al.}}%
            {Interatomic two-electron exchange in layered cuprates}  
\thispagestyle{empty}
\setcounter{page}{1}
\copyrightheading{}
\vspace*{0.88truein}
\fpage{1} 
\centerline{\textbf{REDUCED PAIRING HAMILTONIAN FOR INTERATOMIC}}
\vspace*{0.035truein}
\centerline{\textbf{TWO-ELECTRON EXCHANGE IN LAYERED CUPRATES}}
\vspace*{0.37truein}
\centerline{\footnotesize%
 T.~M.~MISHONOV,$^{\dag,\ddag}$\ \ %
 J.~P.~WALLINGTON,$^{\dag}$\ \ %
 E.~S.~PENEV$^{\ddag}$\ \ {\normalsize and}
 J.~O.~INDEKEU$^{\dag}$
}
\vspace*{0.015truein}
\centerline{$^{\dag}$\footnotesize\it
   Laboratorium voor Vaste-Stoffysica en Magnetisme, %
   Katholieke Universiteit Leuven
}
\baselineskip=10pt
\centerline{\footnotesize\it
   Celestijnenlaan 200 D, B-3001 Leuven, Belgium
}
\vspace*{0.015truein}
\centerline{$^\ddag$\footnotesize\it
   Department of Theoretical Physics, Faculty of Physics,%
   University of Sofia ``St. Kliment Ohridski''
}
\baselineskip=10pt
\centerline{\footnotesize\it
    5 J. Bourchier Blvd., 1164 Sofia, Bulgaria
}
\vspace*{0.225truein}
 
\vspace*{0.21truein}
\abstracts{%
A detailed Linear Combination of Atomic Orbitals (LCAO) tight-binding
model is developed for the layered High-Temperature Superconductor
(HTSC) cuprates. The band structure of these materials is described
using a $\sigma$-band Hamiltonian employing ${\rm Cu}4s$, ${\rm
Cu}3d_{x^2-y^2}$, ${\rm O}2p_{x}$ and ${\rm O}2p_{y}$ atomic orbitals.
The Fermi level and the shape of the resulting Fermi surface are
fitted to recent Angle Resolved Photon Emission Spectroscopy (ARPES)
data to realistically determine the dispersion in the conduction
band. Electron-electron interactions and, ultimately, Cooper pairing
is obtained by introducing a Heitler-London, two-electron exchange
between adjacent orbitals within the ${\rm CuO}_2$ plane.  Finally,
using the LCAO wavefunctions determined by the band structure fit, the
Bardeen-Cooper-Schrieffer (BCS) type kernel is derived for interatomic
exchange.
}{}{}



\vspace*{1pt}\textlineskip
\section{Introduction}
\label{sec:1}
\vspace*{-0.5pt}
\noindent
Since the first discovery in 1986, the high-temperature
superconducting cuprates have attracted great
attention.\cite{OM,DM} There is a wealth of experimental data about
this unusual family of materials which we shall not attempt to review
here.  Rather we will mention only two lines of inquiry: the Angle
Resolved Photoemission Spectroscopy (ARPES) or Angle Resolved
Ultraviolet Photoelectron Spectroscopy (ARUPS) experiments to
determine the normal state band structure, and the ARPES measurements
of the gap in the superconducting state. From these experiments there
are quantitative results for the shape of the Fermi surface and the
angular dependence of the superconducting gap around this surface,
both of which can be compared in detail to any proposed theoretical
model.

The quantity of theoretical work on the cuprates is similarly large,
but can be roughly divided into two camps: one has made considerable
effort to realistically model the band structure of the materials,
from first principles where possible, whilst the other has
concentrated on correlation effects by using idealized ``toy" models
of the Hubbard and $t$-$J$ types.  Whilst both of these camps have made
considerable progress in their own fields, there has been surprisingly
little overlap between the two.  In particular, the band structures
employed in the studies of correlations are rarely more sophisticated
than the inclusion of nearest- or next-nearest-neighbour hoppings on a
square lattice, and bear little resemblance to either the measured or
predicted band structures.

It is the purpose of this paper to analyze microscopic {\em
interatomic} two-electron exchange processes which could be useful to
understand the nature and origin of the superconducting correlations
in the cuprate materials. As it is now almost universally accepted
that it is the physics of the ${\rm CuO}_2$ planes, the common feature
of all the cuprates, which gives the materials such extraordinary
properties our efforts will focus on the properties of these planes
and we will omit the effects of extraneous, material-dependent details
such as the presence of ${\rm CuO}$ chains, orthorhombic distortions,
double planes, etc.  Having devised a model we will fit recent
experimental data to the single-particle sector, so that it accurately
describes the band structure, and then we will proceed to derive an
analytic expression for the BCS-type kernel, or pairing potential,
$V({\bf p},{\bf p}^{\prime})$.  We will leave the solution of the
resulting gap equation, and comparison to experiment, to a later date.

\section{Model}
\vspace*{-0.5pt}
\noindent
In this work we attempt to understand the physics of the ${\rm CuO}_2$
plane using a relatively simple, idealized, tight-binding model,
within the Linear Combination of Atomic Orbitals (LCAO) approximation.
To this end we consider only the four atomic orbitals coming from the
${\rm Cu}3d_{x^2-y^2}$, ${\rm Cu}4s$ and ${\rm O}2p_{\sigma}$ states.
Denoting the positions of the ${\rm Cu}$, ${\rm O}_x$ and ${\rm O}_y$
atoms by ${\bf R}_{\mathrm{Cu}}$, ${\bf R }_x$ and ${\bf R }_y$ respectively,
the in-plane lattice constant by $a_0$ and the unit cell index by
${\bf n}=(n_x,n_y)$, the LCAO wavefunction reads:
\begin{eqnarray}
\psi_{\mathrm{LCAO}}({\bf r})
    = \sum_{{\bf n},\alpha}
        \Big[
            D_{{\bf n}\alpha} \psi_{{\rm Cu}3d_{x^2-y^2}}({\bf r} - {\bf n}a_0 - {\bf R}_{Cu})
            +S_{{\bf n}\alpha} \psi_{{\rm Cu}4s}({\bf r} - {\bf n}a_0 - {\bf R}_{\mathrm{Cu}})
            \\
            +X_{{\bf n}\alpha} \psi_{{\rm O}2p_x}({\bf r} - {\bf n}a_0 - {\bf R}_x)
            +Y_{{\bf n}\alpha} \psi_{{\rm O}2p_y}({\bf r} - {\bf n}a_0 - {\bf R}_y)
        \Big],\nonumber
\end{eqnarray}
where $\Psi_{{\bf n}\alpha}=(D_{{\bf n}\alpha},S_{{\bf
n}\alpha},X_{{\bf n}\alpha},Y_{{\bf n}\alpha})$ is the vector of
amplitudes for the ${\bf n}$th unit cell. With each of these
quantities is associated a Fermion operator satisfying the usual
anticommutation relations $\{\hat{X}^{}_{{\bf
n}\alpha},\hat{X}^{\dagger}_{{\bf m}\beta}\}=\delta_{{\bf
nm}}\delta_{\alpha\beta}$, $\{\hat{X}^{}_{{\bf
n}\alpha},\hat{Y}^{\dagger}_{{\bf m}\beta}\}=0$, etc.  We use these
operators to analyze exchange processes in the ${\rm CuO}_2$ planes.

This model can be conveniently split into two parts; a non-interacting
part describing the single-electron band structure and an interacting
part describing the effects of electron-electron interactions.  We
consider each of these in turn below.

\subsection{Tight-binding band structure}
\vspace*{-0.5pt}
\noindent
When the differential overlap of the orbitals is neglected, the
non-interacting, single-electron Hamiltonian can be parameterized by
three atomic energies, $\epsilon_s$, $\epsilon_d$ and $\epsilon_p$,
and three hopping integrals, $t_{sp}$, $t_{pd}$ and $t_{pp}$,
corresponding to the transitions ${\rm Cu}4s\leftrightarrow {\rm
O}2p_{\sigma}$, ${\rm O}2p_{\sigma}\leftrightarrow {\rm
Cu}3d_{x^2-y^2}$ and ${\rm O}2p_x\leftrightarrow {\rm O}2p_y$
respectively.  In second-quantization notation, the resulting
Bloch-H\"uckel-like Hamiltonian\cite{MP} reads
\begin{eqnarray}
 \hat{H}_{\mathrm{BH}}
    = \sum_{{\bf n}\alpha}
        \bigg\{
            \hat{D}^{\dagger}_{{\bf n}\alpha}
                \Big[
                    \epsilon_d \hat{D}_{{\bf n}\alpha}
                    -t_{pd}
                    \big(
                        -\hat{X}_{{\bf n}\alpha}+\hat{X}_{(n_x-1,n_y)\alpha}
                        -\hat{Y}_{{\bf n}\alpha}+\hat{Y}_{(n_x,n_y-1)\alpha}
                    \big)
                \Big] \\
            +\hat{S}^{\dagger}_{{\bf n}\alpha}
                \Big[
                    \epsilon_s \hat{S}_{{\bf n}\alpha}
                    -t_{sp}
                    \big(
                        -\hat{X}_{{\bf n}\alpha}+\hat{X}_{(n_x-1,n_y)\alpha}
                        -\hat{Y}_{{\bf n}\alpha}+\hat{Y}_{(n_x,n_y-1)\alpha}
                    \big)
                \Big] \nonumber \\
            +\hat{X}^{\dagger}_{{\bf n}\alpha}
                \Big[
                    \epsilon_p \hat{X}_{{\bf n}\alpha}
                    -t_{pd}
                    \big(
                        -\hat{D}_{{\bf n}\alpha}+\hat{D}_{(n_x+1,n_y)\alpha}
                    \big)
                    -t_{sp}
                    \big(
                        -\hat{S}_{{\bf n}\alpha}+\hat{S}_{(n_x+1,n_y)\alpha}
                    \big)
            \nonumber \\ \qquad
                    -t_{pp}
                    \big(
                        \hat{Y}_{{\bf n}\alpha}-\hat{Y}_{(n_x+1,n_y)\alpha}
                        -\hat{Y}_{(n_x,n_y-1)\alpha}+\hat{Y}_{(n_x+1,n_y-1)\alpha}
                    \big)
            \Big] \nonumber \\
            +\hat{Y}^{\dagger}_{{\bf n}\alpha}
                \Big[
                    \epsilon_p \hat{Y}_{{\bf n}\alpha}
                    -t_{pd}
                    \big(
                        -\hat{D}_{{\bf n}\alpha}+\hat{D}_{(n_x,n_y+1)\alpha}
                    \big)
                    -t_{sp}
                    \big(
                        -\hat{S}_{{\bf n}\alpha}+\hat{S}_{(n_x,n_y+1)\alpha}
                    \big)
            \nonumber \\ \qquad
                    -t_{pp}
                    \big(
                        \hat{X}_{{\bf n}\alpha}-\hat{X}_{(n_x-1,n_y)\alpha}
                        -\hat{X}_{(n_x,n_y+1)\alpha}+\hat{X}_{(n_x-1,n_y+1)\alpha}
                    \big)
            \Big]
        \bigg\}. \nonumber
\end{eqnarray}
This Hamiltonian is diagonalized by introducing the Bloch states
$\hat{\Psi}_{{\bf p}\alpha} = $ $ (\hat{D}_{{\bf
p}\alpha},\hat{S}_{{\bf p}\alpha},\hat{X}_{{\bf
p}\alpha},\hat{Y}_{{\bf p}\alpha})$.  The elements of these states are 
defined by the relation
\begin{equation}
\label{bloch_form}
\hat{\Psi}_{{\bf n}\alpha} = \left(
    \begin{array}{c}
        \hat{D}_{{\bf n}\alpha} \\
        \hat{S}_{{\bf n}\alpha} \\
        \hat{X}_{{\bf n}\alpha} \\
        \hat{Y}_{{\bf n}\alpha}
    \end{array}
\right)
    = \frac{1}{\sqrt{N}} \sum_{{\bf p}} e^{i{\bf p}\cdot{\bf n}}
\left(
    \begin{array}{c}
        \hat{D}_{{\bf p}\alpha} \\
        \hat{S}_{{\bf p}\alpha} \\
        e^{i \phi_a} \hat{X}_{{\bf p}\alpha}  \\
        e^{i \phi_b} \hat{Y}_{{\bf p}\alpha}
    \end{array}
\right),
\end{equation}
where $N$ is the number of unit cells, and the two phases are $\phi_a
= \frac{1}{2}(p_x-\pi)$ and $\phi_b = \frac{1}{2}(p_y-\pi)$.  The
Hamiltonian can now be expressed in the more compact form
\begin{eqnarray}
\hat{H}_{\mathrm{BH}}
    = \sum_{{\bf p}\alpha}
        \hat{\Psi}^{\dagger}_{{\bf p}\alpha}
            H_{\bf p}
        \hat{\Psi}^{}_{{\bf p}\alpha},
\end{eqnarray}
where the Hamiltonian matrix is
\begin{equation}
H_{\bf p}
    = \left(
        \begin{array}{cccc}
            \epsilon_d & 0 & t_{pd} s_x & -t_{pd} s_y \\
            0 & \epsilon_s & t_{sp} s_x & t_{sp} s_y \\
            t_{pd} s_x & t_{sp} s_x & \epsilon_p & -t_{pp} s_x s_y \\
            -t_{pd} s_y & t_{sp} s_y & -t_{pp} s_x s_y & \epsilon_p \\
        \end{array}
    \right).
\end{equation}
Here, for the sake of convenience, we have adopted the shorthand
notation of Andersen {\em et al.}:\cite{Andersen1,Andersen2}
$s_x=2\sin(p_x/2)$, $s_y=2\sin(p_y/2)$.

The spectrum of this model is determined by the solutions of the
Heisenberg equation of motion, $i \hbar d_t \hat{\Psi}_{{\bf p}\alpha}
= [\hat{\Psi}_{{\bf p}\alpha},\hat{H}_{\mathrm{BH}}]$, which is equivalent to
solving for the LCAO amplitudes:
\begin{equation}
\left( H_{\bf p} - \epsilon \boldsymbol{1} \right) \Psi_{{\bf p}\alpha}
    =   0.
\end{equation}
This matrix equation is easily solved.  The four eigenvalues are
determined from the secular equation
\begin{equation}\label{det_H}
\det \left( H_{p} - \epsilon 1 \right)
    =   {\cal A}(\epsilon)xy
            + {\cal B}(\epsilon)(x+y)
            + {\cal C}(\epsilon)
    =0
\end{equation}
in which $x=\sin^2(p_x/2)$, $y=\sin^2(p_y/2)$ and the coefficients are
\begin{eqnarray}
{\cal A}(\epsilon)
    &=& 16  \left(
                4t_{pd}^2 t_{sp}^2
                + 2t_{sp}^2 t_{pp} \varepsilon_d
                - 2t_{pd}^2 t_{pp} \varepsilon_s
                - t_{pp}^2 \varepsilon_d \varepsilon_s
            \right),\\
{\cal B}(\epsilon)
    &=& -4\varepsilon_p \left(
                            t_{sp}^2 \varepsilon_d
                            + t_{pd}^2 \varepsilon_s
                        \right),\\
{\cal C}(\epsilon)
    &=& \varepsilon_d \varepsilon_s \varepsilon_p^2.
\end{eqnarray}
In the above, the energies $\varepsilon_d$, $\varepsilon_s$ and
$\varepsilon_p$ are measured relative to their respective atomic
levels: $\varepsilon_d = \epsilon-\epsilon_d$, $\varepsilon_s =
\epsilon-\epsilon_s$ and $\varepsilon_p = \epsilon-\epsilon_p$.

To each of the four solutions of equation (\ref{det_H}) there
corresponds an eigenvector $\Psi_{{\bf p}\alpha}^{B}$ where $B$, the
band index, labels the solutions.  The general (unnormalized) form for
the eigenvector is
\begin{equation}
\Psi_{{\bf p}\alpha}^{B}=
\left(
    \begin{array}{c}
        D^{B}_{\bf p} \\
        S^{B}_{\bf p} \\
        X^{B}_{\bf p} \\
        Y^{B}_{\bf p}
    \end{array}
\right) = \left(
    \begin{array}{c}
        -\varepsilon_s \varepsilon_p^2 + 4 \varepsilon_p t^2_{sp}(x+y)
            - 32 t_{pp} {\tilde t}^2_{sp} xy\\
        -4 \varepsilon_p t_{sp} t_{pd}(x-y) \\
        -(\varepsilon_s \varepsilon_p - 8 {\tilde t}^2_{sp} y) t_{pd} s_x  \\
        (\varepsilon_s \varepsilon_p - 8 {\tilde t}^2_{sp} x) t_{pd} s_y
    \end{array}
\right)
\end{equation}
where ${\tilde t}_{sp}^2 = t_{sp}^2 - \varepsilon_s t_{pp}/2$, and
$\varepsilon_d$, $\varepsilon_s$ and $\varepsilon_p$ are understood to
depend on the appropriate value of $\epsilon$, which in turn depends
upon ${\bf p}$ and the model parameters.  We denote the normalized
eigenvector with a tilde: $\tilde{\Psi}_{{\bf p}\alpha}^{B} =
\Psi_{{\bf p}\alpha}^{B}/|\Psi_{{\bf p}\alpha}^{B}|$.

\subsection{Heitler-London interaction}
\vspace*{-0.5pt}
\noindent
The second-quantized Heitler-London interaction Hamiltonian describes
the interatomic two-electron exchange between the copper and oxygen
orbitals.\cite{MDKP,MGD} It comprises three parts corresponding to
${\rm Cu}4s\leftrightarrow {\rm O}2p_{\sigma}$, ${\rm
O}2p_{\sigma}\leftrightarrow {\rm Cu}3d_{x^2-y^2}$ and ${\rm
O}2p_x\leftrightarrow {\rm O}2p_y$ exchanges with transition
amplitudes $J_{sp}$, $J_{pd}$ and $J_{pp}$ respectively:
\begin{equation}
\hat{H}_{\mathrm{HL}}
    = \hat{H}_{\mathrm{HL}}^{sp} + \hat{H}_{\mathrm{HL}}^{pd} + \hat{H}_{\mathrm{HL}}^{pp},
\end{equation}
where,
\begin{align}
\hat{H}_{\mathrm{HL}}^{sp}
    =& -\frac{J_{sp}}{2} \sum_{\alpha\beta}
        \sum_{{\bf n}}
        \Big[
                \hat{S}^{\dagger}_{{\bf n}\alpha} \hat{X}^{\dagger}_{{\bf n}\beta}
                    \hat{S}^{}_{{\bf n}\beta} \hat{X}^{}_{{\bf n}\alpha}
                +\hat{S}^{\dagger}_{{\bf n}\alpha} \hat{Y}^{\dagger}_{{\bf n}\beta}
                    \hat{S}^{}_{{\bf n}\beta} \hat{Y}^{}_{{\bf n}\alpha}
            \\
                &+\hat{S}^{\dagger}_{(n_x+1,n_y)\alpha} \hat{X}^{\dagger}_{{\bf n}\beta}
                    \hat{S}^{}_{(n_x+1,n_y)\beta} \hat{X}^{}_{{\bf n}\alpha}
                +\hat{S}^{\dagger}_{(n_x,n_y+1)\alpha} \hat{Y}^{\dagger}_{{\bf n}\beta}
                    \hat{S}^{}_{(n_x,n_y+1)\beta} \hat{Y}^{}_{{\bf n}\alpha}
        \Big], \nonumber
\end{align}
\begin{align}
\hat{H}_{\mathrm{HL}}^{pd}
    =& -\frac{J_{pd}}{2} \sum_{\alpha\beta}
        \sum_{{\bf n}}
        \Big[
                \hat{D}^{\dagger}_{{\bf n}\alpha} \hat{X}^{\dagger}_{{\bf n}\beta}
                    \hat{D}^{}_{{\bf n}\beta} \hat{X}^{}_{{\bf n}\alpha}
                +\hat{D}^{\dagger}_{{\bf n}\alpha} \hat{Y}^{\dagger}_{{\bf n}\beta}
                    \hat{D}^{}_{{\bf n}\beta} \hat{Y}^{}_{{\bf n}\alpha}
            \\
                &+\hat{D}^{\dagger}_{(n_x+1,n_y)\alpha} \hat{X}^{\dagger}_{{\bf n}\beta}
                    \hat{D}^{}_{(n_x+1,n_y)\beta} \hat{X}^{}_{{\bf n}\alpha}
                +\hat{D}^{\dagger}_{(n_x,n_y+1)\alpha} \hat{Y}^{\dagger}_{{\bf n}\beta}
                    \hat{D}^{}_{(n_x,n_y+1)\beta} \hat{Y}^{}_{{\bf n}\alpha}
        \Big], \nonumber
\end{align}
\begin{align}
\hat{H}_{\mathrm{HL}}^{pp}
    =& -\frac{J_{pp}}{2} \sum_{\alpha\beta}
        \sum_{{\bf n}}
        \Big[
                \hat{X}^{\dagger}_{{\bf n}\alpha} \hat{Y}^{\dagger}_{{\bf n}\beta}
                    \hat{X}^{}_{{\bf n}\beta} \hat{Y}^{}_{{\bf n}\alpha}
                +\hat{X}^{\dagger}_{{\bf n}\alpha} \hat{Y}^{\dagger}_{(n_x+1,n_y)\beta}
                    \hat{X}^{}_{{\bf n}\beta} \hat{Y}^{}_{(n_x+1,n_y)\alpha}
            \\
                &+\hat{X}^{\dagger}_{(n_x,n_y+1)\alpha} \hat{Y}^{\dagger}_{{\bf n}\beta}
                    \hat{X}^{}_{(n_x,n_y+1)\beta} \hat{Y}^{}_{{\bf n}\alpha}
            \nonumber \\
                &+\hat{X}^{\dagger}_{(n_x,n_y+1)\alpha} \hat{Y}^{\dagger}_{(n_x+1,n_y)\beta}
                    \hat{X}^{}_{(n_x,n_y+1)\beta} \hat{Y}^{}_{(n_x+1,n_y)\alpha}
        \Big]. \nonumber
\end{align}
The complexity of this expression is greatly simplified by moving to
the Bloch basis defined in equation (\ref{bloch_form}).  In this
momentum-space representation it is more natural to label the
fermionic operators by the band index, $B$, than by the orbital
index. Accordingly we separate these indices by writing
\begin{equation}
    \hat{\Psi}_{{\bf p}\alpha}
        = \sum_B \tilde{\Psi}_{{\bf p}\alpha}^B
            \hat{c}_{{\bf p}\alpha}^{B},
\end{equation}
where $\tilde{\Psi}_{{\bf p}\alpha}^B$ is the normalized LCAO
eigenvector coming from the band structure and $\hat{c}_{{\bf
p}\alpha}^{B}$ is the annihilation operator for an electron in the
$B$-band with momentum ${\bf p}$ and spin $\alpha$.  In what follows,
we will drop the band-index and focus solely on the strongly
hybridized $d$-band which contains the Fermi level.

\section{BCS Theory}
\vspace*{-0.5pt}
\noindent
Moving to the Bloch basis and substituting for $\hat{\Psi}_{{\bf
p}\alpha}$ leads to the momentum-space representation of the
Heitler-London interaction term.  At this point we have made no
approximations, aside from neglecting completely filled or empty
bands, and the representation is exact.  However, the resulting
Hamiltonian is too complicated for us to proceed further without
approximations.  The simplest course we can take is to make the
mean-field or BCS approximation whereby fluctuations about the
homogeneous, currentless equilibrium state are ignored. This leads to
a simplified form of the Hamiltonian:
\begin{equation}
\hat{H}_{\mathrm{HL}}
    =   \frac{1}{2N} \sum_{\alpha\beta}
        \sum_{{\bf pp}^{\prime}}
        \hat{c}^{\dagger}_{{\bf p}^{\prime}\beta}
        \hat{c}^{\dagger}_{-{\bf p}^{\prime}\alpha}
        V({\bf p},{\bf p}^{\prime}) \;
        \hat{c}^{}_{-{\bf p}^{}\alpha}
        \hat{c}^{}_{{\bf p}^{}\beta}
\end{equation}
where the BCS pairing-potential is
\begin{align}
V({\bf p},{\bf p}^{\prime})
    =&   \left[
            J_{sp} \tilde{S}_{{\bf p}^{}} \tilde{S}_{{\bf p}^{\prime}}
            + J_{pd} \tilde{D}_{{\bf p}^{}} \tilde{D}_{{\bf p}^{\prime}}
        \right]
                \left[
            W_{xx^{\prime}}
            \tilde{X}_{{\bf p}^{}} \tilde{X}_{{\bf p}^{\prime}}
            + W_{yy^{\prime}}
            \tilde{Y}_{{\bf p}^{}} \tilde{Y}_{{\bf p}^{\prime}}
        \right]\\
    & -   \frac{J_{pp}}{2}
        \tilde{X}_{{\bf p}^{}} \tilde{X}_{{\bf p}^{\prime}}
        W_{xx^{\prime}} W_{yy^{\prime}}
        \tilde{Y}_{{\bf p}^{}} \tilde{Y}_{{\bf p}^{\prime}}
        \nonumber.
\end{align}
Here we have introduced a new notation:
\begin{align}
W_{xx^{\prime}}
    &= 
      c_x^{} c_x^{\prime} - s_x^{} s_x^{\prime}
    =   4\cos\left(\frac{p_x+p_x^{\prime}}{2}\right),\\
W_{yy^{\prime}}
    &= c_y^{} c_y^{\prime} - s_y^{} s_y^{\prime}
    =   4\cos\left(\frac{p_y+p_y^{\prime}}{2}\right).
\end{align}
Note that $W_{XX^{\prime}}$ and $W_{YY^{\prime}}$ are sums of {\em
separable} terms, which is to say that the $p_x$ and $p_x^{\prime}$
dependencies can be factored out.  Thus $V({\bf p},{\bf p}^{\prime})$
is also a sum of separable terms.

For a sufficiently strong, attractive potential, and a low enough
temperature, the BCS Hamiltonian gives rise to a superconducting
condensate of Cooper pairs with zero center-of-mass momentum.
Following the BCS prescription we obtain the following self-consistent
expression for the superconducting gap:\cite{Tinkham}
\begin{equation}
\label{eqn:gapequation}
\Delta({\bf p})
    = -\frac{1}{N}\sum_{{\bf p}^{\prime}}
        V({\bf p},{\bf p}^{\prime})
        \frac{\tanh\left(\frac{E({\bf p}^{\prime})}{2T}\right)}{2E({\bf p}^{\prime})}
        \Delta({\bf p}^{\prime}),
\end{equation}
where the quasiparticle energies are $E({\bf p})=\sqrt{\xi^2({\bf
p})+|\Delta({\bf p})|^2}$, and $\xi({\bf p})=\epsilon({\bf p})-\mu$ is
the normal state energy in the $d$-band relative to the chemical
potential or Fermi energy.

\section{Discussion}
\vspace*{-0.5pt}
\noindent
In this paper we derived the matrix elements of different interatomic
two-electron exchange processes and obtained the corresponding gap
equation. The analysis demonstrates that the $B_{1g}$ symmetry of the
order parameter,\cite{TK} required for the description of the ARPES
and Josephson data for the gap, cannot be obtained by considering only
interatomic interactions. We thus arrive at the conclusion that other
exchange processes should be found and taken into account to explain
the pairing interaction in the cuprates, if we wish to follow the
traditional BCS scheme.

%
\nonumsection{References}


\begin{thebibliography}{99}

\bibitem{OM} J.~Orenstein and A.~J.~Millis, \textit{Science} {\bf 288}, 468
(2000).

\bibitem{DM} ``\textit{High-Tc Superconductors and Related
Materials}'', Editors S.-L.~Drechsler and T.~Mishonov, pp.~487--504,
Kluwer Academic Publishers, Dordrecht, 2001.

\bibitem{MP}
T.~Mishonov and E.~Penev, \textit{J.~Phys.: Condens. Matter} {\bf 12}, 143
(2000).

\bibitem{Andersen1} O.~K.~Andersen, O.~Jepsen, A.~I.~Liechtenstein,
and I.~I.~Mazin, \textit{Phys. Rev.} {\bf B49}, 4145 (1994); O.~K.~Andersen,
S.~Y.~Savrasov, O.~Jepsen, and A.~I.~Liechtenstein, \textit{J. Low Temp. Phys.}
{\bf 105}, 285 (1996), and references therein.

\bibitem{Andersen2} O.~K.~Andersen, A.~I.~Liechtenstein, O.~Jepsen,
and F.~Paulsen, \textit{J.~Phys.~Chem.~Solids} {\bf 56}, 1573 (1995).

\bibitem{MDKP} T.~M.~Mishonov, A.~A.~Donkov, R.~K.~Koleva, and
E.~S.~Penev, \textit{Bulgarian J.~Phys.} {\bf 24}, 114 (1997). 

\bibitem{MGD} T.~M.~Mishonov, A.~V.~Groshev, and A.~A.~Donkov,
\textit{Bulgarian J. Phys.} {\bf 25}, 62 (1998).

\bibitem{Tinkham} See, for example, M.~Tinkham, ``\textit{Introduction
to superconductivity}'' (McGraw-Hill, New York, 1996), or
J.~B.~Ketterson and S.~N.~Song, ``\textit{Superconductivity}''
(Cambridge University Press, Cambridge 1999).

\bibitem{TK} C.~C.~Tsuei and J.~R.~Kirtley, \textit{Rev. Mod. Phys.} {\bf 72}
969 (2000).
\end{thebibliography}
\end{document}